\documentstyle[prl,aps]{revtex}
\begin{document}

\title{Comment on "Temperature dependence of the Raman-active B$_{1g}$
and A$_{1g}$ modes in YNi$_2$B$_2$C" }
\author {A. P. Litvinchuk}
\address{Texas Center for Superconductivity, University of Houston,
	Houston, TX 77204-5932, USA}
\date{February 2, 1998}
\maketitle

\begin{abstract}
Both principal conclusions of Hirata and Takeya paper 
(Phys. Rev. B~{\bf 57}, 2671 (1998)) dealing with 
an analysis of the electron-phonon coupling strengths and stating the 
existence of two boron-related fully symmetric phonon modes in 
YNi$_2$B$_2$C are argued not to be supported by experimental data 
and not justified.
\end{abstract}


In a recent paper \onlinecite{hira98}, Hirata and Takeya argued that
"no strong electron-phonon coupling is operative in superconducting
YNi$_2$B$_2$C" in contradiction with other experimental and theoretical
data (see, e.g., Ref. \onlinecite{yanson97} and references cited therein).
This statements is based on the fact that
"no renormalization effects for the Ni B$_{1g}$ mode" have been
detected in Ref. \onlinecite{hira98}. Another conclusion of the paper refers to
the existence of two components of the high frequency boron
A$_{1g}$ mode due to the presence of $^{11}$B and $^{10}$B isotopes.
I believe that both principal conclusions are not supported by the
presented experimental data and not justified. It is important, however,
to address the question on the strengths of the electron-phonon coupling
in superconductors, especially in view of recent spectroscopic findings
of extremely strong electron-phonon coupling effects\cite{hadj97}
in newly synthesized Ba$_2$Ca$_{n-1}$Cu$_n$O$_x$  high temperature
superconductor\cite{chu97}.

The arguments are as follows:

(i)  The authors of Ref. \onlinecite{hira98} did not take into account heating 
of the strongly
absorbing metallic sample by intense (50 $mW$) laser radiation. Earlier
experiments on YNi$_2$B$_2$C indicate \cite{my95} that heating effects
become substantial for power densities above 100 $W/cm^2$. Because
of heating, which is determined not only by the power density, but also
the laser spot and sample geometry and heat conductivity of the
sample\cite{maks}, at the lowest "nominal cryostat" temperature 10~K the real
temperature  within the scattered volume could exceed T$_c$ = 14.2~K.

(ii) Even in the case there is no "substantial" heating effects, 
Hirata and Takeya have taken 
{\it only one} experimental measurement below the superconducting
transition temperature T$_c$ (Fig. 4 of Ref. \onlinecite{hira98}). 
This makes the discussion about 
possible superconductivity induced phonon self-energy effects simply {\it pointless};

In general, there is a big concern about the reliability of the data presented
in Fig. 4 of Ref. 1, which is due to an unacceptable quality of fits to the
experimental data in Fig. 3, especially for temperatures above 40K.
Moreover, the only fitted temperature dependence of the B$_{1g}$ phonon 
parameters (linewidth $2\gamma$) to anharmonic decay model is
incorrect. Expression 
$2\gamma (\omega, T) = 2\gamma (\omega, 0) [1+2n(\omega /2, T)] + const $, 
where $\omega $ and $n$ are the phonon frequency and Bose-Einstein occupation 
number, respectively, suggests weak temperature dependence of $2\gamma$ 
at low temperatures and its linear increase (but not saturation as shown
in Fig. 3) for $kT \gg \hbar \omega /2 $.

As far as boron-related A$_{1g}$ mode is concerned\cite{my95,victor},
one has to point out that the existence of two boron isotopes does not
necessarily cause appearance of two corresponding Raman lines.
This depends largely\cite{barker} on the phonon dispersion relations
(phonon density of states) in YNi$_2$B$_2$C.
For semiconducting Ge, for example, the mass
ratio of the heaviest and the lightest isotopes is comparable
with those for $^{11}$B and $^{10}$B. In isotopically mixed
$^{70}$Ge/$^{76}$Ge crystal, however, one observes experimentally 
only {\it one line} in the Raman spectrum\cite{fuchs,zhan98}.

There are a number of other facts which put the conclusion of the paper
\onlinecite{hira98} under question:

(i)  Following the  $^{10}$B/$^{11}$B isotope distribution of 20/80,
one might expect, in the first approximation, the intensity of the
line with lower frequency to be higher by a factor of four, 
as it is the case for, e.g., Si:B single crystals\cite{martin}.
This contradicts with experimental data of Fig. 2 (Ref. 1), where
two high frequency lines in the range 800-860 cm$^{-1}$ 
are comparable in intensity;

(ii)  Further, for closely situated vibrational modes of the same
origin and symmetry one expects a {\it similar} anharmonicity-related 
frequency shift upon variation of temperature. This is again in 
strong contradiction with the data of Fig. 2, where two lines in question 
exhibit {\it distinct} behavior and harden by 10 and 28 cm$^{-1}$ 
upon cooling from room temperature to 10K;

(iii) The room temperature spectrum (Fig. 2, Ref. 1) of
YNi$_2$B$_2$C exhibits a structure around 960 cm$^{-1}$ which is
neither fitted or even discussed in the paper.
It seems that the signal-to-noise
ratio of the spectra presented in Figs. 1 and 2 of Ref. 1 is not
sufficient to draw confident conclusions about the spectral distribution
of the Raman scattering intensity in the frequency range where boron
A$_{1g}$ vibration occurs. It is possible that the lower frequency
component around 800 cm$^{-1}$ is due to Raman scattering from an
impurity phase or luminescence.


\end{document}